\documentstyle[aps,prl,twocolumn,amsmath,epsf,floats]{revtex}

\newcommand{\bb}{{\bf b}}

\newcommand{\bA}{{\bf A}}
\newcommand{\bq}{{\bf q}}

\newcommand{\bd}{{\bf d}}
\newcommand{\bx}{{\bf x}}

\newcommand{\br}{{\bf r}}
\newcommand{\bomega}{\boldsymbol {\omega}}

\begin{document}
\draft

\twocolumn[\hsize\textwidth\columnwidth\hsize\csname @twocolumnfalse\endcsname
\title{Superpenetration of a high energy $Q \bar Q$ bound state \\
through random color fields}
\author{H.~Fujii and T.~Matsui}
\address{Institute of Physics, University of Tokyo
3-8-1 Komaba, Meguro, Tokyo 153-8902 }
\date{revised:\today}
\maketitle
\begin{abstract}
The transmission amplitude of a color dipole through a random external
color field is computed in the eikonal approximation in order to study
the absorption of high energy quarkonium by nuclear target. It is
shown that the internal color state of the dipole becomes randomized
and all possible color states are eventually equi-partitioned, while
the probability of finding a color singlet bound state attenuates not
exponentially, but inversely proportional to the distance $L$ of the
random field zone which the dipole has traveled. 

\end{abstract}
]                                       

\noindent 1.
The suppression of the quarkonium production in nucleus-nucleus ($AB$)
collisions has been extensively studied both experimentally and
theoretically\cite{WJ98} since it has been proposed as a signal of the
QCD plasma formation.\cite{MS86} The pattern of the observed $J/\psi$
suppression in  light ion induced reactions ($AB$ with $B \ll
A$)\cite{NA38} as well as in $pA$ collisions\cite{pA} have been well
reproduced by the nuclear absorption model.\cite{GH92,KLNS97} In this
model the ratio of the observed production cross section $\sigma_{AB}$
in $AB$ collisions to the one scaled from the elementary $pp$
collisions $\sigma_{pp}$ is given by the simple formula, 
\begin{equation}\label{exp}
\frac{\sigma_{AB}}{AB \sigma_{pp}} =  e^{-\sigma_{\rm abs} n_0 L}
                                   =  e^{- L/L_{\rm abs}}   ,
\end{equation}
where $L$ is the effective length of the nuclear medium the formative
$J/\psi$ (or $c \bar c$) travels through, $\sigma_{\rm abs } \sim$
6--7 mb is the absorption cross section of the "$J/\psi$" due to the
collision with individual nucleon, and $n_0$ = 0.16~fm$^{-3}$ is the
mean nucleon density in nucleus. Further suppression observed in the
Pb-Pb experiments at CERN away from this "base-line" has been
investigated as an anomaly possibly related to the plasma
formation. \cite{BO96,HS00,WJ98}

The above exponential form of the $L$-dependence is obtained if one
assumes a classical stochastic process for the multiple collisions of
$c \bar c$ pair in nuclear matter. One should however note that this
treatment is valid only when the characteristic time of individual
collision is shorter than the time interval between the successive
collisions. It is known that the coherence in the multiple collisions
changes the high energy asymptotic behavior of the scattering
qualitatively.\cite{HF83,BI96}

Each collision of a $c \bar c$ dipole with nucleon have a
characteristic time scale over which the quantum coherence of the
wavefunction is important: 
in the rest frame of the pair it may be
given by $\tau_c \sim 1/\Delta E \sim$ 0.3 fm with binding energy 
$\Delta E$, while for moving pairs
one should take into account the Lorentz time dilatation effect. 
In the Pb-Pb experiment at CERN-SPS, the Lorentz time dilatation factor
of the $J/\psi$ produced in the midrapidity region is $\gamma \sim 10$
in the nuclear rest frame, hence the $c \bar c$ pair propagates over
$t_c = \tau_c \gamma \sim$ 3 fm typically in the target rest frame
which is of the same order of magnitude with the mean internucleon
distance in nuclei ($d = 2 ~(3 / 4 \pi n_0)^{1/3} = $ 2 fm). In the
RHIC (LHC) experiments, for which $\gamma = $ 100 (2000) this
coherence time becomes 30 (600) fm which is considerably larger than
the mean nucleon distance and become even greater than the size of the
nucleus. 
This problem has been pointed out recently also in \cite{KTH01}.

This simple estimate suggests that the independent stochastic
collision picture is only marginally satisfied at SPS energy and
should break down at higher collider energies where the coherence
effect would become essential; it is more natural to consider that the
multiple interactions with nucleons along its path take place {\it 
  simultaneously}. It is the purpose of the present work to 
investigate how $L$-dependence of the nuclear absorption is modified 
at collider energies from the naive form (1) due to the quantum coherence. 
For simplicity, here we will not go into the discussion of the production
mechanism of the $Q \bar Q$ pair in the nuclear collision which involves
color octet, as well as color singlet,  initial states as produced by 
fusion or fragmentation of incident partons (gluons) in nuclear 
target\cite{KS96,HP99,KTH01}; we will
instead consider a much simpler problem how a high energy 
(color singlet) $Q \bar Q$ bound state injected in nuclear target will be 
absorbed by multiple collision in nucleus.  

\vskip 10pt 

\noindent 2.
Similar problem has been studied previously for the absorption of high
energy positronium ($e^+e^-$) while passing through metallic
foil. \cite{LN81,LP81} It has been found that when the positronium
energy becomes ultrarelativistic $E \gg 2m_e$ the positronium flux
attenuates not by the exponential law but by the power law $\sim
L^{-1}$. This phenomenon was called 
{\it  superpenetration}.\cite{LN81}

It is easy to see how this happens.\cite{LP81} 
The transmission amplitude of the positronium through metallic foil
may be computed by the eikonal approximation assuming that the
individual members of the positronium travel along a straight line
trajectory in random atomic electric fields. The eikonal phase factor
for a particle with charge $e$ passing at impact parameter ${\bf b}$
under the influence of the atomic electric field described by the
electric potential $\phi ({\bf x})$ is given by 
$ e /v \int d z \phi ({\bf b}, z) $
hence the transmission amplitude of the high energy electron-positron
pair ($v \sim 1$), passing at 
${\bf b}_1= {\bf b} + {\bf d}/2$ and at
${\bf b}_2= {\bf b} - {\bf d}/2$
is simply 
\begin{eqnarray}
U_{e^+ e^-} ({\bf b}, {\bf d} ) & = & 
e^{  - i e \int d z \phi ({\bf b}_1 , z)  
     + i e \int d z \phi ({\bf b}_2 , z) } 
\simeq
e^{ - i {\bf q} \cdot {\bf d}} ,
\end{eqnarray}
where  ${\bf d} = {\bf b}_1 - {\bf b}_2$
is the transverse size of the dipole and 
${\bf q} = e \int dz \nabla_{\bot} \phi ( {\bf b}, z ) 
= - e \int dz {\bf E}_\bot ({\bf b}, z )$ 
is the net relative transverse momentum acquired by the dipole under
the action of the transverse electric field ${\bf E}_\bot $. In this
case the net momentum transfered to the dipole vanishes due to the
charge neutrality of the dipole.

Since the configuration of the electric field seen by the pair is
random, the relative transfered momentum $\bq = (q_1, q_2)$ should be
taken as a random variable which is to be averaged over with a
normalized distribution $f( {\bf q} ; L )$; viewing the process as
diffusion in the transverse momentum space, we expect a simple
two-dimensional gaussian distribution 

\begin{equation}
 f ( {\bf q} ; L ) d{\bf q} 
=
 \frac{1}{\pi \langle q^2 \rangle } 
 e^{ - \bq^2 / \langle q^2 \rangle} d\bq
\end{equation}
with the variance increasing linearly with $L$:
$\langle q^2 \rangle = c n_A L$
where $n_A$ is the atomic density and $c$ is a dimensionless
constant. The penetration probability of the positronium state
$|\varphi_0 \rangle$ is thus given by 
\begin{eqnarray}\label{PL}
P_{\varphi_0} (L)  
& = &
\overline{ | \langle \varphi_0 | U ({\bf q}) |  \varphi_0 \rangle |^2 }
\nonumber \\ 
& = & \int d \bq f ( {\bf q};L ) | F ( {\bf q}) |^2 \   ,
\end{eqnarray}
where 
\begin{equation}\label{ff}
  F ( {\bf q} ) =  
  \int d{\bf  r}_\bot d z 
  | \varphi_0 ( {\bf r}_\bot, z ) |^2  
  e^{ - i  {\bf q} \cdot {\bf r}_{\bot} } 
\end{equation}
is the transverse form factor of the positronium state. 
Inserting (\ref{ff}) into (\ref{PL}) and then performing the integral
over the transfered momentum $\bq$, we obtain an alternative
expression for $P_{\varphi_0} (L)$:
\begin{equation}\label{PL'}
P_{\varphi_0}  (L) 
=
 \int d \br_\bot d \br'_\bot \rho_0 (\br_\bot)  \rho_0 (\br'_\bot) 
e^{ - \frac{1}{4} \langle q^2 \rangle 
      \left( \br_\bot - \br'_\bot \right)^2  } 
\ , 
\end{equation} 
where 
$ \rho_0 (\br_\bot) \equiv 
  \int d z | \varphi_0 ( {\bf r}_\bot, z ) |^2 $ 
is the probability distribution of the dipole size ${\bf r}_\bot$ 
in the bound state $\varphi_0 ({\bf r}_\bot, z )$.

For a thin target with small $\langle q^2 \rangle$, we can expand 
the exponential in (\ref{PL'}) and find 
\begin{eqnarray}\label{smallL}
  P_{\varphi_0}  (L) 
\simeq
  1 - \frac{1}{2}  \langle {\bf r}_\bot^2 \rangle \langle q^2 \rangle
= 
  1 - \frac{1}{2}  \langle {\bf r}_\bot^2 \rangle c n_A L  
\ .
\end{eqnarray}
This result can be obtained by the first Born approximation to the
eikonal transmission amplitude and coincides with the small $L$
expansion behavior of the exponential decay form (1), if we identify
$ c \langle {\bf r}_\bot^2 \rangle = 2 \sigma $.

However, for large value of $L$, $P_{\varphi_0} (L)$
deviates from the exponential
form and exhibits  power law behavior,
\begin{equation}\label{largeL}
  P_{\varphi_0}  (L) 
\simeq
  \frac{1}{\pi \langle q^2 \rangle} \int d \bq | F ( {\bf q} ) |^2
= \frac{\Delta}{c n_A L/2}  
= \frac{\Delta \langle {\bf r}_\bot^2 \rangle}{\sigma n_A L}  
\ ,
\end{equation}
where 
$\Delta = (1/2 \pi) \int d \bq | F ( {\bf q} ) |^2 
        = 2 \pi \int d \br_\bot \rho_0^2 (\br_\bot)$.  
Thus the probability that the positronium penetrates through the 
metallic foil of thickness $L$ attenuates inversely proportional to
$L$, not in the exponential form as in (1). We observe from the
formula (\ref{PL}) that this $L^{-1}$ dependence originates from the
depletion of the central value of the distribution $f ( 0 ; L )$ due
to the diffusion of transfered momentum in the phase space as $L$
increase. The formula (\ref{PL'}) indicates also that this is caused
by the destructive interference of the transmission amplitudes for
different dipole sizes with 
$| \br_\bot - \br'_\bot | > 1/\sqrt{\langle q^2 \rangle}$.

Similar asymptotic behavior in multiple collision has been noted also 
by several authors in analyses of high energy hadronic process 
in nuclear target. \cite{ZKL81,HPG90,GB96}
In particular we note the work of 
H\"{u}fner {\it et al.} \cite{HPG90} who treated the problem of 
the nuclear absorption of a quarkonium in an abelian model of
random nuclear color fields with a very similar analysis as 
presented above for the positronium case and found the $1/L$ 
asymptotic behavior for  the bound state survival probability.  
However, their conclusion was plagued  by the result of \cite{HLN90} 
in which the authors claim the stochastic behavior reappears 
when one includes the non-abelian color degrees of freedom.

The $1/L$-dependence was rediscovered recently by Baym
et al.\cite{GB96} by consideration of the Lorentz time-dilatation
effect for the quantum fluctuation of the transverse  size of the
projectile together with color neutrality argument\cite{BBGG81,BM88},
similar to the treatment in \cite{LN81} for positronium absorption.
In their semi-classical treatment, our formula (\ref{PL'}) is
replaced in effect by 
\begin{equation}
  P_{\varphi_0}^{\rm ~s.c.} (L) 
=
  \int d \br_\bot \rho_0 (\br_\bot)  
  e^{ - \sigma (\br_\bot) n_A L  } 
\ , 
\end{equation}
which also gives the $1/L$ asymptotic behavior if the effective cross
section of the dipole of size $\br_\bot$  goes as $\sigma (\br_\bot)
\propto \br^2_\bot$.\footnote{We note here that in this semi-classical 
treatment $1/L$
  arises because only small dipole components of  the wavefunction can
  penetrate through the medium due to their small absorption cross
  section $ \sigma (\br_\bot)$.  This is not the case in our full
  quantum treatment where the dipoles of large size also contributes
  to the penetration probability.}
Although the quantum coherence between successive collisions was
seemingly ignored in these treatments,  their result suggests that
the $L^{-1}$ attenuation law is rather universal not specific to the
abelian  nature of the positronium problem.  In what follows, we
extend the works of \cite{LP81} to include non-abelian  nature of
the color interaction and show that the survival probability of the
$c \bar c$ bound state will indeed attenuate by the $L^{-1}$ law,
rather than exponential, at large $L$.

\vskip 10pt 

\noindent 3.
The new features which arises in the case of non-Abelian color
interaction is associated with the color degrees of freedom in the
internal wavefunction of the pair. In the SU(3), there are eight
different color orientations for the octet $c \bar c$ states in
addition to the color singlet configuration.  Action of the external
color field will cause transitions between these 9 different color
states.\footnote{Action of the external field on the color octet 
$c \bar c$
  state may also lead to gluon emission from the dipole. We ignore
  this gluon radiation in the following discussion and leave this
  effect for future study.}

To study these qualitatively new aspects, we consider here the SU(2)
version of the color interaction. The external gauge fields are then given by  
$\bA_\mu ( x )= \sum A^a_\mu( x ) \tau^a/2$ 
where $\tau^a$ ($a = 1, 2, 3$) are the ($2 \times 2$)
Pauli matrices and the internal color states of the dipole are labeled
by the isospin quantum number: for simplicity, we denotes the singlet
state by $| 0 \rangle  \equiv | 0, 0 \rangle$, and three iso-triplet
states by
 $( | 1 \rangle, | 2 \rangle, | 3 \rangle)  \equiv 
( \frac{1}{\sqrt{2}}( | 1, 1 \rangle + i | 1, - 1 \rangle),
  \frac{1}{\sqrt{2}}( | 1, 1 \rangle - i | 1, - 1 \rangle),
  | 1, 0 \rangle ) $; 
we use Roman letters $a, b, \cdots$ for labeling  the three triplet
states and Greek letters, $\alpha, \beta, \cdots$ for all four states
including the singlet state.

The eikonal transmission amplitude of the color dipole $Q \bar Q$ may
be written as  
\begin{eqnarray}\label{Uq}
  U_{Q \bar Q} ( {\bf b}, {\bf d} ) 
& = & 
  {\cal P} e^{ i g \int d x^+ \bA_{1}^- ({\bf b}_1, x^+ )
             + i g \int d x^+ \bA_{2}^- ({\bf b}_2, x^+ ) }
\nonumber \\
& \simeq &
  {\cal P} e^{ i { \bomega} ({\bf b}) \cdot \bd + i \theta ({\bf b}) } 
\ ,
\end{eqnarray}
where we have used the light-cone coordinates\footnote{Here we work 
on the target rest frame where the pair is propagating along  the 
trajectory $x_- = 0$.}
 $x^\pm = (t \pm z)/\sqrt{2}$
with $A^\pm =(A^0 \pm A^z)/\sqrt{2}$, the path-ordered product along
the integration paths on the light-cone ($x^- = 0$) is implied by
${\cal P}$ and  
\begin{eqnarray}
  {\bf \bomega} ({\bf x}_\bot) 
& \equiv & 
\nabla_\bot \int d x^+ g A^{a-} ({\bf x}_\bot, x^+) (\tau^a_1 - \tau^a_2)/4
\ ,
\\
  \theta ({\bf x}_\bot) 
& \equiv & 
\int  d x^+ gA^{a-}  ({\bf x}_\bot, x^+ ) ( \tau^a_1 + \tau^a_2)/2
\  .
\end{eqnarray}
Here $\tau_1$ ($\tau_2$) acts on internal isospin (color) space of $Q$
($\bar Q$) respectively and $\bomega = ( \omega_1, \omega_2 )$ and
$\theta$ are ($4 \times 4 $) matrices which operate on the internal
isospin (color) space of the dipole.

Some qualitative insights about how non-Abelian color interaction
works may be obtained by expanding this amplitude by the Born series: 
\begin{equation}
U_{Q \bar Q} ( {\bf b}, {\bf d} ) 
=  
1 + i {\bomega} ({\bf b}) \cdot \bd + i \theta ({\bf b}) +\cdots \ ,
\end{equation}
where we have shown only the first Born term explicitly. 
The color matrix $\bomega$ has non-vanishing matrix elements only 
between the singlet state $| 0 \rangle$ and the triplet states 
$| a \rangle$ :
\begin{eqnarray}
\langle a  | U_{Q \bar Q} (\bb, \bd ) | 0  \rangle 
& =  &
i \langle a | {\bomega} ({\bf b}) | 0 \rangle \cdot \bd + \cdots 
\nonumber \\
& = &
 i \frac{\bd}{2} \cdot \nabla_\bb \int d x^+ g A^{a-} ({\bf b}, x^+) 
+ \cdots  
\end{eqnarray}
while the matrix $\theta$ gives non-vanishing expectation values only
between the color triplet states:
\begin{eqnarray}
\langle a  | U_{Q \bar Q} (\bb, \bd ) | b \rangle
 & = &
\delta_{ab} + i \langle a | \theta ({\bf b}) | b \rangle + \cdots
\nonumber \\
& = & 
\delta_{ab} + i \epsilon_{abc} \int d x^+ g A^{c-} ({\bf b}, x^+) 
+ \cdots \ ,
\nonumber \\
\end{eqnarray}
where the antisymmetric tensor $\epsilon_{abc}$ is the structure
constant of the SU(2) group.

Imagine that a color singlet dipole with frozen transverse size $\bd$
is injected into a random color field.  It will then be transformed to
color triplet state by the action of the field and therefore the
probability that it remains in color singlet state, $P_{0,0}$, will
attenuate with the distance $L$ as it travels through the random
field. In the first Born approximation, it may be given by 
\begin{eqnarray}\label{P00}
P_{0, 0} (\bd ; L) 
& \simeq &
 1 - \sum_a \overline{ 
       | \langle a | {\bomega} ({\bf b}) | 0 \rangle \cdot \bd |^2 } 
\  ,
\end{eqnarray}
where the average is taken over the random field, or equivalently, we
may consider the matrix ${\bomega}$ as a random variable. Since
$\overline{\nabla_i A^a (\bx) \nabla_j A^b (\bx') } 
= \delta_{ab}\delta_{ij}  f (\bx - \bx')$ 
by symmetry, we may write
\begin{equation}\label{alpha2}
\overline{ \langle a | \omega_i  | 0 \rangle 
           \langle 0 | \omega_j  | b \rangle } 
=  \delta_{ij} \delta_{ab} \gamma_0 L 
\  ,
\end{equation}
where $\gamma_0$ is a constant of the dimension of [volume]$^{-1}$.
\footnote{The value of $\gamma_0$ is a measure of the transverse
correlation of the color electric fields in the target rest frame and we 
expect $\gamma_0 \sim g^2 \Lambda_{QCD}^3$ on dimensional
ground.  In partonic picture, it may be related 
to the saturation scale $Q_s$ of the transverse parton momentum 
distribution\cite{M01} in a boosted frame. 
Here we take this constant as a pure phenomenological parameter
which may be determined by comparing our result (\ref{thin}) to data.}
Inserting this into (\ref{P00}), we obtain
\begin{equation}\label{leading}
P_{0, 0} (\bd; L) \simeq 1 - 3 \gamma_0 L {\bf d}^2 \ .
\end{equation}
This result coincides with the leading term of the expansion of the
naive exponential form (\ref{exp}) if we interpret $\sigma_{\bf d} = 3
\gamma_0 {\bf d}^2 / n_0 $ as the absorption cross section of the
dipole of size $\bf d$.  The proportionality of the absorption cross
section to the square of the dipole size is the manifestation of the
color transparency.\cite{BBGG81,BM88} 

If the injected color singlet state is described by the internal
wavefunction $| \varphi_0 \rangle$, we should replace ${\bf d}^2$ by
its expectation value $\langle {\bf d}^2 \rangle_{\varphi_0}$: 
this procedure may be justified by multiplying 
$\langle a  | {\bomega} ({\bf b}) | 0 \rangle $ by  
$\int d \bd d z \varphi_n^*  (\bd, z ) \varphi_0 (\bd, z ) $
and use the closure sum
$\sum \varphi_n^* (\bd,  z) \varphi_n (\bd', z' ) 
= \delta (\bd - \bd') \delta (z - z')$ 
over all intermediate states $n$.
It then follows that if we interpret 
$\sigma_{\varphi_0} = 3 \gamma_0 \langle {\bf d}^2 \rangle_{\varphi_0} / n_0$ 
as the absorption cross section of the bound state, our result
coincides with the exponential form (\ref{exp}) which is based on the
classical stochastic assumption for the multiple scattering
process. We will come back to this result later.

This is a rather surprising result since in our derivation we have
taken the coherence of scatterings from different part of the field
fully into account. It is the random field average in (\ref{alpha2})
which eliminates the interference terms of these summed amplitude. The
formula (\ref{leading}), however, is based on the first Born
approximation and is valid only when $\sigma n_0 L \ll 1$.

\vskip 10pt 

\noindent 4.
For large $L$, we divide the volume occupied by the random field into
$n$ different small zones along the trajectory of the dipole in the
same spirit as in \cite{HLN90,HW98}.   Each zone may be considered as
belonging to each individual hadron (nucleon) in the
nucleus\cite{HLN90} or one could even divide the color field in the
same hadron into smaller portions of cells further\cite{HW98}.  What
is assumed here is that the fields belonging to the different zones
are uncorrelated and therefore one can take average over the field in
each zone independently.  The size of the zone $l$ characterizes the
coherence of the fluctuating color field in the nucleus or in the
hadron. The random color field has been also introduced as a model of
the Weiz\"{a}cker-Williams fields for small $x$ parton distribution in
\cite{MV94}. 

\begin{figure}[tb]
\begin{center}
\epsfxsize=0.25\textwidth
\epsffile{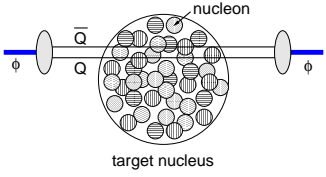}
  \caption{Schematic picture of color dipole propagating in nucleus
    which is viewed as a source  of random field. }
\label{fig:1}
\end{center}
\end{figure}

The eikonal amplitude is now factorized into the product of the
transmission amplitude through each zone:
\begin{equation}
U_{Q \bar Q} (\bd,\bb)  = U_n \cdots U_2 U_1 
\  ,
\end{equation}
where the transmission amplitude for the $i$th zone, $U_i$, is given
by the formula similar to (\ref{Uq}): 
\begin{equation}
U_i   \simeq 
e^{i {\bomega}^{(i)} ({\bf b}) \cdot \bd + i \theta^{(i)} ({\bf b}) }
\ .
\end{equation}
Let us introduce the matrix 
\begin{equation}\label{calm}
{\cal M}^{(i)}_{\alpha, \alpha' ; \beta, \beta'} (\bd, \bd') 
= 
\overline{ \langle \alpha | U_i (\bd) | \beta \rangle 
           \langle \beta' | U_i^\dagger (\bd') | \alpha' \rangle }
\  ,
\end{equation}
where the average over the random field in the $i$th zone is implied
by the overline. Then the survival probability of the color singlet
state $| \varphi_0 \rangle$ after penetrating through $n$ 
zones of uncorrelated random color fields is given by
\begin{eqnarray}\label{survive}
S_{\varphi_0} (n) 
& = & 
\overline{| \langle \varphi_0 | U_{Q \bar Q} |  \varphi_0 \rangle |^2 }
\nonumber \\ 
& = & 
\int d \bd d \bd'  \rho_0 (\bd) \rho_0 (\bd') K_{0,0} (\bd, \bd'; n)
\  ,
\end{eqnarray}
where $\rho_0 (\bd) = \int dz |\varphi_0 (\bd, z) |^2$ is the
probability distribution of the color singlet dipole of size $\bd$ in
the wavefunction $\varphi_0 (\bd, z)$\footnote{
Here we take non-relativistic limit of the light-cone
wavefunction and ignore the mixing of the higher Fock space 
components which would give energy dependence of
the dipole distribution.} 
and the kernel of integration is
expressed in term of $\cal M$ as 
\begin{eqnarray}\label{K00}
&&  K_{0,0} (\bd, \bd'; n)
\nonumber \\
 & = &
 \overline{\langle 0 | U_{Q \bar Q} (\bd, n)   | 0 \rangle 
           \langle 0 | U_{Q \bar Q}^\dagger (\bd', n)   | 0 \rangle}
\nonumber \\
& = & \sum 
  {\cal M}_{0,0 ; \alpha_{n-1}, \alpha_{n-1}' } ^{(n)} 
 \cdots 
  {\cal M}_{ \alpha_2, \alpha_2' ; \alpha_1, \alpha_1' }^{(2)}
  {\cal M}_{ \alpha_1, \alpha_1' ; 0 , 0}^{(1)} 
\ ,
\end{eqnarray}
where the sum is taken over all intermediate color states 
($\alpha_i, \alpha_i'= 0, 1, 2, 3$) . 
The formula (\ref{survive})  is non-abelian extension of the formula 
(\ref{PL'}) for the positronium penetration probability. 
It still remains to compute the kernel function $K_{0,0} (\bd, \bd'; n)$.

For a sufficiently small zone the matrix ${\cal M}_i $ may be
evaluated by the perturbation theory by expanding
\begin{eqnarray*}
U_i ( {\bf b}, {\bf d} ) 
& =  & 
1 + i {\bomega}^{(i)} ({\bf b}) \cdot \bd + i \theta^{(i)} ({\bf b})  \\
&& - \frac{1}{2} \left( {\bomega}^{(i)} ({\bf b}) \cdot \bd +
      \theta^{(i)} ({\bf b}) \right)^2  + \cdots \  .
\end{eqnarray*}
Calculation of the matrix elements of ${\cal M}_i $ requires the color
average of the matrix elements of the products of $\bomega^{(i)}$ and
$\theta^{(i)}$ in addition to (\ref{alpha2}) (with $L$ replaced by
$l$). Since there is no preferred color direction and fields in the
different zones are uncorrelated, we have 
\begin{equation}\label{alphabeta}
\overline{ \langle a | \bomega^{(i)} | b  \rangle 
           \langle b'| \theta^{(j)} | a' \rangle } = 0
\end{equation}
while  
\begin{equation}\label{beta2}
\overline{ \langle a | \theta^{(i)} | b  \rangle 
           \langle b'| \theta^{(j)} | a' \rangle } 
= 
\epsilon_{abc} \epsilon_{a'b'c}\delta_{ij} \eta ( {\bf b}) l 
\ ,
\end{equation}
where $\eta$ is a constant with the dimension of [length]$^{-1}$
which is also related to the transverse correlation length of the
gauge fields as $\gamma_0$ and $\eta \sim g^2 \Lambda_{QCD}$, 
$l$ is the thickness of each zone. We also need
\begin{eqnarray}\label{alpha20}
\overline{ \langle a | \bomega^{(i)} \cdot \bomega^{(j)} | b  \rangle }
& =  &  \delta_{a b}  \delta_{ij} \gamma_0 l 
\ ,
\nonumber \\
\overline{ \langle a | \theta^{(i)}\theta^{(j)} | b  \rangle } 
& = &  \delta_{a b} \delta_{ij} \eta l  
\end{eqnarray}
to compute all matrix elements of the $16 \times 16$ matrix 
${\cal M} (\bd,\bd')$ to the order $g^2$.

Due to the superselection rules encoded in (\ref{alpha2}),
(\ref{alphabeta}), (\ref{beta2}) and (\ref{alpha20}), the matrix
${\cal M} (\bd,\bd')$ becomes block diagonal when it is expressed in
the bases of the eigenstates of the total isospin and its $z$
components. Hereafter we adopt this new representation where Greek
letters $\alpha$, $\beta$, $\dots$, stand for 
$(T, T_z) = (0, 0), (1,1), (1, 0), (1, -1) $. 
In particular, the block matrix for 
$\alpha = \alpha'$ and $\beta = \beta'$:  
\begin{equation}\label{M}
M_{\alpha, \beta} (\bd, \bd')  
= {\cal M}^{(i)}_{\alpha, \alpha; \beta, \beta } (\bd, \bd') 
\end{equation}
is given explicitly by 
\begin{eqnarray}\label{Mab}
 M (\bd, \bd' )  =  
\left( 
\begin{array}{cccc}
1 -  3 r &  r'     &  r'   &     r'       \\
r'     & 1 -r - s  &  s    &     0        \\
r'       &  s   & 1 - r -  2s &  s        \\
r'       &  0      &  s    & 1 - r - s    
\end{array}
\right) 
\nonumber \\
&&\hspace*{-1cm}
\end{eqnarray}
with $r =  \frac{1}{2} \gamma_0  l  ( \bd^2 + {\bd'}^2 ) $, 
$r' =  \gamma_0  l  \bd \cdot \bd'$ and 
$s = \eta ( {\bf b}) l $
where each column from top to bottom and each row from left to right
now corresponds to the isospin states, $(0, 0), (1, 1), (1, 0), 
(1,-1)$, respectively. All matrix elements with $\alpha = \alpha'$
and $\beta \neq \beta'$ which couple the block matrix 
$M_{\alpha, \beta}(\bd, \bd')$ to the other components of the matrix
${\cal M}^{(i)}$ vanish by the superselection rules and 
$\delta_{\alpha\beta}
\delta_{\alpha' \beta'} = 0$. We can express $K_{0,0} (\bd,\bd' ; n)$
in terms of the the eigenvalues $\lambda_i$ and the corresponding
(normalized) eigenvectors ${\bf v}^{(i)}$ of the matrix $M (\bd,\bd')$
as: 
\begin{eqnarray}\label{K00A}
K_{0,0} (\bd,\bd' ; n) & =  & \sum_i \lambda_i^n ( v_1^{(i)} )^2 
= \sum_i \lambda_i^n f_{0 ; i} \ ,
\end{eqnarray}
where $f_{0; i} \equiv ( v_1^{(i)} )^2 $ is the fraction of the color singlet component in the $i$th eigenstate.
We find from the expression (\ref{Mab}) of $M (\bd,\bd')$ only two eigenstates have non-zero mixing of
color singlet components for which
\begin{eqnarray}\label{lambda}
 \lambda_1 & =& 
1 - \frac{1}{2} 
\left(  d_+^2 + d_-^2 -  
        \sqrt{d_+^4 + d_-^4 - d_+^2  d_-^2} \right) \gamma_0 l \ ,
\nonumber \\
\lambda_2 & =& 
1 - \frac{1}{2} 
\left(  d_+^2 + d_-^2 +  
        \sqrt{d_+^4 + d_-^4 - d_+^2  d_-^2} \right) \gamma_0 l
\end{eqnarray}
and 
\begin{eqnarray}\label{fraction}
f_{0; 1} & = & 
\frac{1}{2} - \frac{d_+^2 + d_-^2}
                   {4 \sqrt{d_+^4 + d_-^4 - d_+^2 d_-^2} }
 \ ,
\nonumber \\
f_{0; 2} & = & 
\frac{1}{2} + \frac{d_+^2 + d_-^2}
                   {4 \sqrt{d_+^4 + d_-^4 - d_+^2  d_-^2} } 
\ ,
\end{eqnarray}
where $d_\pm^2 = (\bd \pm \bd' )^2$.  
Inserting  (\ref{lambda}) and (\ref{fraction}) into (\ref{K00}),
we obtain
\begin{eqnarray}\label{K00'}
K_{0,0} (\bd,\bd' ; n) 
& = &
 \frac{\lambda_1^n + \lambda_2^n}{2} - 
 \frac{(d_+^2 + d_-^2) ( \lambda_1^n - \lambda_2^n )}
      {4 \sqrt{d_+^4 + d_-^4 - d_+^2  d_-^2}}
\ .
\end{eqnarray}

The diagonal components of the kernel $K_{0,0} (\bd,\bd;n)$ can be
interpreted by the construction as the probability that a color
singlet rigid dipole of size $\bd$ is found as color singlet after
going through $n$ uncorrelated random color zones. Similarly, at 
$\bd = \bd'$ the matrix $M (\bd, \bd')$ reduces to a matrix 
$P (\bd) = M(\bd, \bd)$ 
whose matrix element 
$P_{\alpha, \beta} (\bd) = | \langle \alpha | U_i | \beta \rangle |^2$, 
can be interpreted as the probability of the dipole of the initial
color state $\alpha$ being found in the color state $\beta$ in the
final state after going through the $i$th random field zone. Since $P$
is a symmetric stochastic matrix $P = ( p_{\alpha \beta} )$ with 
$ \sum_\alpha p_{\alpha \beta} = \sum_\beta p_{\alpha \beta} = 1$, 
it has an eigenvalue of 1 for the eigenvector 
${\bf v} =  ( 1, 1, 1, 1)^T$ and other eigenvalues are smaller than 1.
Moreover, the matrix $P^n$ will approach with increasing $n$ to the
matrix with all elements equal to $1/4$.
This latter property implies that the transition probability from the
initial isospin (color) singlet state to all color states are equal;
namely, there is equi-partitioning of all different isospin states in
the final state. 
The physical reason for the appearance of the
stochastic color equi-partitioning is the cancellation of all
interference terms due to the random color averaging.  

The approach to the color equilibrium can be calculated explicitly
from (\ref{K00'}) as 
\begin{eqnarray}
K_{0,0} (\bd,\bd ; n)
& =  & 
   \frac{1}{4} 1^n+ \frac{3}{4} (1- 4 r )^n
 = \frac{1}{4} + \frac{3}{4} e^{ - n \kappa}
\end{eqnarray}
where 
$\kappa = - \ln ( 1 - 4 r (\bd,\bd) ) \simeq 4r = 4 \gamma_0 l  \bd^2 $.

This stochastic behavior of the motion in the internal color space was
already noticed in \cite{HLN90,HW98}. It resembles to the exponential
decay if one ignores the asymptotic value $1/4$ ($1/9$ in the case of
SU(3)). This is not the entire story, however: the relaxation process 
in the color space is limited by the finite number of color degrees of freedom
and the bound state survival probability still attenuates by the $1/L$ law
due to the other continuous degrees of freedom, namely the transverse
size $\bd$ of the color dipole which is frozen during the collision, 
as we shall show now. 

The survival probability of the bound state can be obtained by inserting (\ref{K00'})
into (\ref{survive}) and performing the integral over $\bd$ and
$\bd'$. It is important here to take into account the interferences of
the transmission amplitudes at different dipole size $\bd \neq \bd'$
as we have seen in the calculation of the positronium penetration
probability: the information about the  internal momentum transfer
$\bq$ by the scattering in the random field is contained in the
dependence on $\bd - \bd'$ of the kernel $K_{0,0} (\bd, \bd'; n)$,
namely in its off-diagonal components,

We first check that our formula contains the thin target result by
setting $n=1$: 
\begin{eqnarray}\label{thin}
S_{\varphi_0} (1) 
& = & \int d \bd d \bd'  \rho_0 ( \bd)  \rho_0 ( \bd') 
\left[ 1 - \frac{3}{4} (d_+^2 + d_-^2) \gamma_0 l \right] 
\nonumber \\
& = & 1 - 3 \langle \bd^2 \rangle \gamma_0 L  = 1 - L/L_{\rm abs} \ .
\end{eqnarray}
where we have used $L=l$ and 
$L_{\rm abs} = 1/3 \langle \bd^2 \rangle \gamma_0$ may be interpreted 
as the absorption length of the bound state.
This is exactly what we have obtained already. 

For thick targets with large $n$, the terms with the largest
eigenvalue, $\lambda_1$, dominate so that 
\begin{eqnarray}
S_{\varphi_0} (n) 
& \simeq & \int d \bd d \bd' \rho_0 ( \bd) \rho_0 ( \bd') 
\nonumber \\
& & \quad \times
\left[ \frac{1}{2} -
 \frac{(d_+^2 + d_-^2) }{4 \sqrt{d_+^4 + d_-^4 - d_+^2  d_-^2}}
\right] \lambda_1^n 
\ .
\end{eqnarray}
To estimate the remaining integrals we observe that 
the integrand is symmetric by the interchange 
$d_+ \leftrightarrow d_-$, or equivalently by 
$\bd' \leftrightarrow - \bd'$, and the factor
\begin{equation}
\lambda_1^n = e^{n \ln \lambda_1} \simeq 
e^{- \frac{1}{2} 
\left( d_+^2 + d_-^2 - 
       \sqrt{d_+^4 + d_-^4 - d_+^2  d_-^2} \right) \gamma_0 n l}
\end{equation}
becomes 1 when $\bd = \pm \bd'$ but decreases very rapidly as
$\bd \pm \bd'$ increases.
Expanding the exponent in terms of $d_-^2 = (\bd - \bd')^2$ and
setting $\bd = \bd'$ in the prefactor, we may estimate roughly as
\begin{equation}\label{thick}
S_{\varphi_0} (n)  \simeq  
2 \int d \bd d \bd'  \rho_0^2 ( \bd) 
 \frac{1}{4}  e^{- \frac{3}{4} (\bd -\bd')^2 \gamma_0 L}
 = \frac{L_0}{L} \ ,
\end{equation}
where the constant $L_0$ is given by 
\begin{equation}
L_0 = 
2 \times \frac{1}{4} \times 2 \times \Delta \langle \bd^2 \rangle  
L_{\rm abs} 
\end{equation}
with $\Delta = 2 \pi \int d \bd \rho_0^2 ( \bd) $ and we have used 
$3 \langle \bd^2 \rangle \gamma_0 = \sigma n_0$. 
This behavior is essentially the same as we have seen in the
positronium case, (\ref{largeL}), except for the factors $2 \times
\frac{1}{4} \times 2$:
the first factor 2 arises from the symmetry by 
$d_+ \leftrightarrow d_-$ which is specific to the case of SU(2) color
charge; the second factor $\frac{1}{4}$ originates from the
equi-partitioning in the color degrees of freedom and will be replaced
by $1/N_c^2 = 1/9$ for SU(3) color; finally the last factor 2 has
arisen because only one of the two eigenvalues $\lambda_1$ and
$\lambda_2$ contributes at large $L$.  In the case of SU(2) color
charge these factors cancel accidentally, while in the case of SU(3)
color charge, they are replaced by 
$1 \times \frac{1}{3^2} \times 2 = \frac{2}{9}$.\cite{HF02}

\begin{figure}[tb]
\begin{center}
\epsfxsize=0.4\textwidth
\epsffile{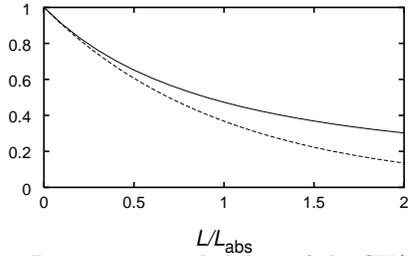}
\caption{Penetration probability of the SU(2) color singlet state as a
  function of the target thickness $L/L_{\rm abs}$.  The exponential
  form (dashed curve) is shown for reference.
\label{fig2}}
\end{center}
\end{figure}

Having established the asymptotic behavior of the survival
 probability $S_{\varphi_0} (L)$ we examine how the transition from
 the thin target result (\ref{thin}) to the thick target asymptotic
 behavior  (\ref{thick}) takes place using the gaussian normalized
 dipole size distribution: 
 $\rho_0 (\bd ) = \frac{1}{\pi d_0^2} e^{- \bd^2/d_0^2} $
which may be obtained from the ground state wave function of the
 non-relativistic heavy quark of the reduced particle mass $\mu$ in
 the harmonic oscillator potential with the frequency $\omega = 1/\mu
 d_0^2$.  
In this case, $\Delta \langle \bd^2 \rangle = 1$ and two length scales
 $L_{\rm abs}$ and $L_0$ coincide.  The survival probability
 $S_{\varphi_0}$ then becomes a function of a single dimensionless
 variable $x = L/L_{\rm abs}$.  
The numerical result is plotted in Fig. \ref{fig2} and compared with
 the exponential decay form obtained by the exponentiation of the thin
 target result (\ref{thin}):  
$ 1 - L/ L_{\rm abs} \to e^{-L/L_{\rm abs}}$. 
It is seen that the exponential absorption formula gives an
overestimate of 20 \% for the suppression at $L = L_{\rm abs}$ in the
 case of SU(2) color charge, while this is slightly reduced in the
 case of SU(3) color charge as shown in Fig. 3.

In summary, we have studied the absorption of a high energy $Q \bar Q$
bound state passing through random color fields taking fully into
account the quantum coherence of the multiple scatterings while
assuming the transverse size of the dipole is frozen by the Lorentz time
dilatation.  It was
shown that the absorption is weaker than that given by the exponential
damping form commonly used in phenomenological models for nuclear
absorption and is given instead by the power law inversely
proportional to the thickness of the target asymptotically.  Although 
our calculation of penetration probability of a $Q \bar Q$ bound state 
in nuclear target is not directly applicable to the quarkonium production problem 
in nuclear collision, our result suggests that a special caution is needed 
to use the naive 
nuclear absorption model of quarkonium suppression at high energies,
especially at RHIC and LHC energies.  However, it remains to
be seen how the coherence effect will show up in the nuclear
collisions when one includes the production mechanism of 
$Q \bar Q$ pair\cite{HP99}.   We are now working on the problem 
and the result will be reported elsewhere.

\vskip 10pt

\noindent{\bf Acknowledgments}:
We like to thank Gordon Baym, Arthur Hebecker, Dima Kharzeev, Larry
McLerran, and Raju Venugopalan for discussions at various stages of
the present work.  One of the authors (H.F.) thanks the hospitality of
RBRC at BNL, where a part of this work was done.  This work is supported
in part by the Grants-in-Aid for Scientific Research of the Ministry
of Education and Sciences (Monkasho) of Japan (No.\ 1340067).  

\begin{figure}[tb]
\begin{center}
\epsfxsize=0.4\textwidth
\epsffile{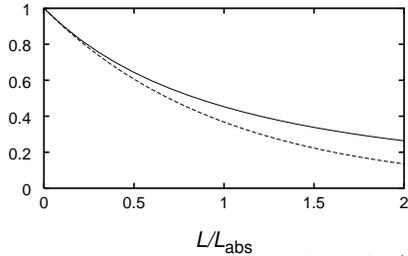}
\caption{Penetration probability of the SU(3) color singlet state as a
function of the target thickness $L/L_{\rm abs}$.
\label{fig3}}
\end{center}
\end{figure}

\end{document}